\documentclass[12pt]{article}
\usepackage{amsmath,amssymb,bm,epsfig}
\usepackage{graphicx}
\usepackage{dcolumn}
\usepackage{color}
\usepackage{longtable}
\newcolumntype{d}[1]{D{.}{.}{#1}}

\renewcommand{\thefootnote}{\fnsymbol{footnote}}
\def\slash#1{\not\!\!#1}

\begin{document}

\title{
\begin{flushright}
\begin{minipage}{0.2\linewidth}
\normalsize
EPHOU-19-006\\
KEK-TH-2118 \\*[50pt]
\end{minipage}
\end{flushright}
{\Large \bf 
Wavefunctions and Yukawa Couplings on Resolutions of $T^2/\mathbb{Z}_N$ Orbifolds
\\*[20pt]}}

\author{Tatsuo~Kobayashi$^{1}$\footnote{
E-mail address:  kobayashi@particle.sci.hokudai.ac.jp}, \ \ 
Hajime~Otsuka$^{2}$\footnote{
E-mail address: hotsuka@post.kek.jp
}, \ and \
Hikaru~Uchida$^{1}$\footnote{
E-mail address: h-uchida@particle.sci.hokudai.ac.jp}
\\*[20pt]
$^1${\it \normalsize 
Department of Physics, Hokkaido University, Sapporo 060-0810, Japan} \\
$^2${\it \normalsize 
KEK Theory Center, Institute of Particle and Nuclear Studies, KEK,}\\
{\it \normalsize 1-1 Oho, Tsukuba, Ibaraki 305-0801, Japan} \\*[50pt]}

\date{
\centerline{\small \bf Abstract}
\begin{minipage}{0.9\linewidth}
\medskip 
\medskip 
\small
We propose matter wavefunctions on resolutions of $T^2/\mathbb{Z}_N$ singularities with constant magnetic fluxes. 
In the blow-down limit, the obtained wavefunctions of chiral zero-modes result in those on the magnetized $T^2/\mathbb{Z}_N$ orbifold models, but the wavefunctions of $\mathbb{Z}_N$-invariant zero-modes receive the blow-up effects around fixed points of $T^2/\mathbb{Z}_N$ orbifolds. 
Such blow-up effects change the selection rules and  Yukawa couplings among the chiral zero-modes as well as 
the modular symmetry, 
in contrast to those on the magnetized $T^2/\mathbb{Z}_N$ orbifold models. 
\end{minipage}
}

\begin{titlepage}
\maketitle
\thispagestyle{empty}
\end{titlepage}

\renewcommand{\thefootnote}{\arabic{footnote}}
\setcounter{footnote}{0}

\tableofcontents

\section{Introduction} 
Toroidal orbifold models~\cite{Dixon:1985jw,Dixon:1986jc}, originally constructed in the heterotic $E_8\times E_8$ or $SO(32)$ string theories~\cite{Gross:1985fr,Gross:1985rr}, 
are theoretically and phenomenologically interesting, since all the couplings are calculable not only within the framework of conformal 
field theory (CFT), but also in the quantum field theory. 
Indeed, in the four-dimensional low-energy effective action of superstring theory on toroidal orbifold backgrounds, 
Yukawa couplings and higher-order couplings among matter fields are determined by 
stringy CFT calculations \cite{Hamidi:1986vh,Dixon:1986qv,Burwick:1990tu,Erler:1992gt,Choi:2007nb} and 
integrals of their internal wavefunctions in field-theoretical calculations. 
Furthermore, toroidal orbifold backgrounds with magnetic fluxes and Wilson lines provide us with 
phenomenologically interesting model building including possibility of realizing quark and lepton masses and 
mixing angles.(See, e.g., Refs.~\cite{Abe:2008sx,Abe:2014vza,Abe:2016jsb}.)

To go beyond the toroidal orbifold models, blow-up models are another useful model building approach, where 
the singularities are replaced by the Eguchi-Hanson spaces~\cite{Eguchi:1978xp}.\footnote{Smooth Calabi-Yau compactifications with gauge background are other useful model building approaches~\cite{Candelas:1985en}, but most of the couplings are understood at the topological level. (See for the recent work, e.g., Refs.~\cite{Otsuka:2018oyf,Otsuka:2018rki}.)} 
However, it is difficult to use the CFT technique on resolutions of general toroidal orbifolds. 
In the field-theoretical approach, metric and gauge fluxes are 
explicitly constructed in a certain class of resolution of toroidal orbifolds like $\mathbb{C}^N/\mathbb{Z}_N$ 
with $N\geq 2$~\cite{Nibbelink:2007rd} and stringy corrections are discussed on them~\cite{Leung:2019oln}. 
The topological quantities can be also derived by employing the toric geometry~\cite{Nibbelink:2007pn}, 
but the matter wavefunction and their couplings are not fully explored so far.

In this paper, we propose the wavefunctions of chiral zero-modes on 
the blow-ups of $T^2/\mathbb{Z}_N$ orbifolds, where the orbifold fixed points are replaced by a part of a sphere. 
Then, we aim to  explicitly calculate Yukawa couplings on 
the blow-ups of  $T^2/\mathbb{Z}_N$ orbifolds.
Our analysis is applicable to  more general factorisable tori like $T^2\times T^2\times T^2/\mathbb{Z}_N$. 
We find that in the blow-down limit, the obtained matter wavefunctions approach to the well-known wavefunctions on toroidal orbifolds. 
The Yukawa couplings among chiral zero-modes as well as the modular symmetry are different from  
the toroidal orbifold results, 
because $\mathbb{Z}_N$-invariant zero-modes receive the blow-up effects around fixed points of $T^2/\mathbb{Z}_N$ orbifolds.

The remainder of this paper is organized as follows. 
In Sec.~\ref{sec:2}, after briefly reviewing the matter wavefunctions of torus and toroidal orbifolds, 
we consider the blow-ups of several toroidal orbifolds, where the orbifold fixed points are replaced by a part of $S^2$. 
As a consequence of the blow-up effects, zero-mode wavefunctions are changed to the wavefunctions on $S^2$. 
Then, we discuss the conditions to smoothly connect the wavefunctions on toroidal orbifolds and $S^2$. 
The normalization of wavefunctions are shown in the end of Sec.~\ref{sec:2}. 
From the obtained deformed wavefunctions, we calculate the overlap integrals of wavefunctions, namely Yukawa couplings 
as shown in Sec.~\ref{sec:3}. The modular symmetry is  also discussed. 
Finally, Sec.~\ref{sec:con} is devoted to the conclusion. 
In Appendix \ref{app}, we show detailed calculations on the wavefunction normalization and Yukawa couplings.

\section{Zero-mode wavefunctions on resolutions of $T^2/\mathbb{Z}_N$}
\label{sec:2}
\subsection{Wavefunctions on $T^2$ and $T^2/\mathbb{Z}_N$}
\label{subsec:2_1}
In this section, we first briefly review the wavefunctions of chiral zero-modes, starting from 6-dimensional supersymmetric Yang-Mills theory on 
toroidal background with $U(1)$ magnetic fluxes~\cite{Cremades:2004wa}. 
The background flat metric in the complex coordinates $z=x+\tau y$ is chosen as
\begin{align}
g=
(2\pi R)^2 
\begin{pmatrix}
0 & \frac{1}{2} \\
\frac{1}{2} & 0 \\
\end{pmatrix}
,
\end{align}
where $R$ and $\tau$ denote the radius and complex structure of the torus, respectively. 
The $U(1)$ magnetic flux satisfying the Hermitian Yang-Mills equation is given by
\begin{align}
F=\frac{iM}{2{\rm Im}\tau} dz\wedge d\bar{z}, 
\end{align}
which must be quantized on $T^2$, namely $(2\pi)^{-1}\int_{T^2}F=M\in \mathbb{Z}$. The above magnetic flux is derived from the following vector 
potential
\begin{align}
A=\frac{M}{2{\rm Im}\tau} {\rm Im}(\bar{z}dz). 
\end{align}

On this gauge background, the zero-mode equation of fermion on $T^2$
\begin{align}
\Psi(z,\bar{z}) = 
\begin{pmatrix}
\psi_+\\
\psi_-\\
\end{pmatrix}
\end{align}
originating from the internal wavefunction of gaugino, is given by
\begin{align}
\slash{D}\Psi=0. 
\end{align}
Since the magnetic flux $M$ generates the net-number of chirality through the index theorem on $T^2$, 
either component $\psi_+$ or $\psi_-$ has a solution of zero-mode equation. In particular, when $M$ is positive (negative), 
$\psi_+$ ($\psi_-$) has the $|M|$ number of degenerate zero-modes. 
In terms of the Jacobi theta function
\begin{align}
\vartheta
\begin{bmatrix}
a\\
b
\end{bmatrix}
(z, \tau)
=
\sum_{l\in \mathbb{Z}}
e^{\pi i (a+l)^2\tau}
e^{2\pi i (a+l)(z+b)}
, 
\end{align}
the zero-mode solution is known to be
\begin{align}
\psi_+^{j,M}(z)&=\left(\frac{2M{\rm Im}\tau}{{\cal A}^2}\right)^{1/4} e^{i\pi Mz{\rm Im}(z)/{\rm Im}\tau} 
\vartheta
\begin{bmatrix}
\frac{j}{M}\\
0
\end{bmatrix}
(Mz, M\tau), \qquad (M>0),
\nonumber\\
\psi_-^{j,|M|}(z)&=\left(\frac{2|M|{\rm Im}\tau}{{\cal A}^2}\right)^{1/4} e^{i\pi |M|z{\rm Im}(z)/{\rm Im}\tau} 
\vartheta
\begin{bmatrix}
\frac{j}{|M|}\\
0
\end{bmatrix}
(|M|z, |M|\tau), \qquad (M<0),
\label{eq:T2wavefcn}
\end{align}
where $j=0,1,...,(|M|-1)$ represents the zero-mode index 
 and ${\cal A}=4\pi^2 R^2 {\rm Im}\tau$ is the 
area of torus determined by the orthonormality condition. 
Here and in what follows, Wilson lines are not included in our analysis for simplicity. 
The lowest-mode solution of scalar fields $\phi^{j,M}(z)$, originating from the internal component of the 6-dimensional vector field, is given by the same functional form with massless fermions. 
Although supersymmetry is broken by the background magnetic flux, 
we can realize four-dimensional supersymmetric vacua 
in higher-dimensional supersymmetric Yang-Mills theory on factorizable tori 
such as 10-dimensions by choosing magnetic fluxes in a proper way.
In such models, lowest scalar modes become massless. 
Throughout this paper, we assume that supersymmetry is preserved in the whole 
system and we focus on one of the tori.

Next, we move on to the wavefunctions of chiral zero-modes on toroidal orbifolds $T^2/\mathbb{Z}_N$. 
In the simple $T^2/\mathbb{Z}_2$ orbifold case, the torus $T^2\simeq \mathbb{C}/\Lambda$ with $\Lambda$ being 
a 2-dimensional lattice, is further identified with $\mathbb{Z}_2$-transformation $z\rightarrow -z$, under which 
there exist four-fixed points. 
Thus, the zero-mode solutions are categorized by two classes, $\mathbb{Z}_2$-even and -odd zero-modes, namely~\cite{Abe:2008fi}
\begin{align}
\psi_{T^2/\mathbb{Z}_2^\pm}^{j,M} = 
\left\{
\begin{array}{c}
\frac{1}{\sqrt{2}} e^{i\pi M z {\rm Im}(z)/{\rm Im}(\tau)} 
\left(
\vartheta
\begin{bmatrix}
\frac{j}{M}\\
0
\end{bmatrix}
(Mz, M\tau) \pm 
\vartheta
\begin{bmatrix}
\frac{M-j}{M}\\
0
\end{bmatrix}
(Mz, M\tau)
\right) \qquad (0<j<\frac{M}{2})
\\
e^{i\pi M z {\rm Im}(z)/{\rm Im}(\tau)} 
\vartheta
\begin{bmatrix}
\frac{j}{M}\\
0
\end{bmatrix}
(Mz, M\tau) \qquad (j=0,\frac{M}{2})
\end{array}
\right.
,
\label{eq:T2Z2}
\end{align}
up to a normalization factor which is explicitly shown later. 
It is noted that the zero-mode wavefunctions satisfy $\vartheta^{j,M}(-z)=\vartheta^{M-j, M}(z)$ 
and $\vartheta^{j,M}(z)$ with $j=0, M/2$ are the $\mathbb{Z}_2$-even wavefunctions. 
In a similar way, the wavefunctions on $T^2/\mathbb{Z}_N$ orbifolds are described by 
\cite{Abe:2013bca,Abe:2014noa,Kobayashi:2017dyu}
\begin{align}
\psi_{T^2/\mathbb{Z}_N^m}^{j,M} = \frac{1}{\sqrt{N}} \sum_{k=0}^{N-1} (\rho_m)^k \psi_{T^2}^{j,M}(\rho^kz),
\label{eq:T2ZNfunc}
\end{align}
with $\rho_m = e^{2\pi i m/N}$ ($m\in \mathbb{Z}$), satisfying $\rho_m^N=1$.
(See for more explicit forms Ref.~\cite{Kobayashi:2017dyu}.) 
Here, $\psi_{T^2}^{j,M}$ is the wavefunction on $T^2$ as shown in Eq.~(\ref{eq:T2wavefcn}) and 
the bosonic wavefunction $\phi^{j,M}_{T^2/\mathbb{Z}_N}$ is the same with the fermionic one.

\subsection{Blow-ups of $T^2/\mathbb{Z}_N$}
\label{subsec:2_2}
In this section, we cut out the $T^2/\mathbb{Z}_N$ orbifold singularities and replace it by a part of $S^2$. 
The reason why we use $S^2$ is understood from the discussion of the Euler number of $T^2/\mathbb{Z}_N$~\cite{Walton:1987bu}. 
For example, in the case of $T^2/\mathbb{Z}_2$ with four fixed points, the Euler number on $T^2/\mathbb{Z}_2$ removing four fixed points is given by
\begin{align}
\frac{\chi(T_2)-4}{2}=-2,
\end{align}
where we use $\chi(T^2)=0$, $\chi=1$ for a point and an order of $\mathbb{Z}_2$ is 2. 
After replacing each fixed point with the disk, we obtain the Euler number of $T^2/\mathbb{Z}_2$
\begin{align}
\frac{\chi(T_2)-4}{2}+4=2,
\end{align}
which is equivalent to the Euler number of $S^2$. 
The above discussion is applicable to $T^2/\mathbb{Z}_N$ case. 
In the following analysis, we replace fixed points of $T^2/\mathbb{Z}_N$ with a part of sphere and discuss the wavefunction 
around the blow-up region of $T^2/\mathbb{Z}_N$.  
Note that the $T^2/\mathbb{Z}_4$ orbifold includes $\mathbb{C}/\mathbb{Z}_2$ singularity  in addition to 
$\mathbb{C}/\mathbb{Z}_4$ singularities and the $T^2/\mathbb{Z}_6$ orbifold includes $\mathbb{C}/\mathbb{Z}_2$ and 
$\mathbb{C}/\mathbb{Z}_3 $ singularities  in addition to 
$\mathbb{C}/\mathbb{Z}_6$ singularity, although the origin $(x,y)=(0,0)$ on  $T^2/\mathbb{Z}_N$ always corresponds to 
the $\mathbb{C}/\mathbb{Z}_N$ singularity.

\subsubsection{$T^2/\mathbb{Z}_2$}
\label{subsubsec:T2Z2}
\begin{figure}[htbp]
\centering
  \begin{minipage}{0.3\columnwidth}
  \centering
     \includegraphics[scale=0.6]{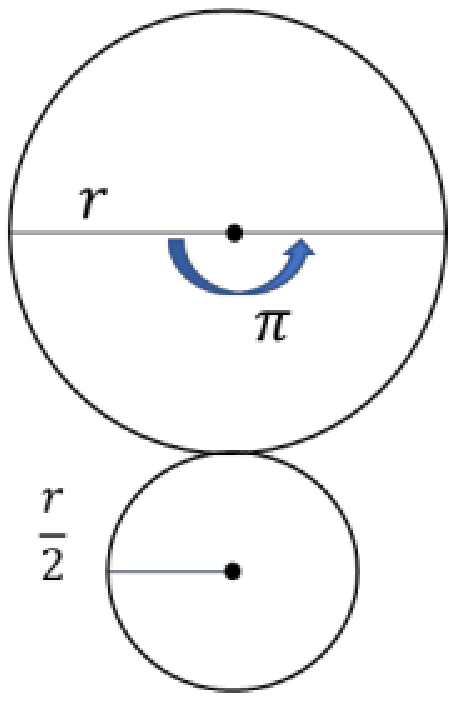}
   \end{minipage}
  \begin{minipage}{0.3\columnwidth}
   \centering
    \includegraphics[scale=0.6]{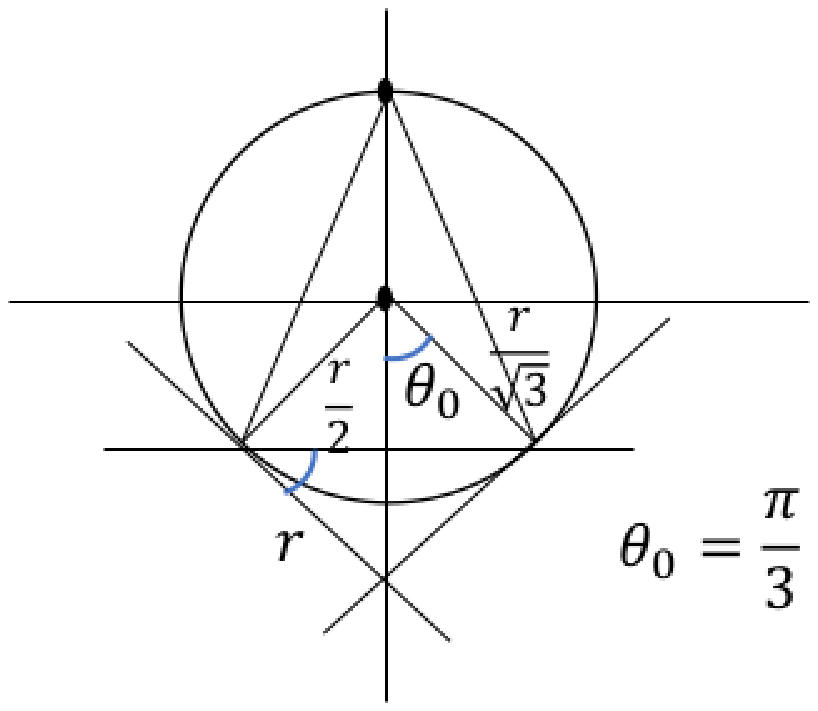}
  \end{minipage}
    \caption{The left panel shows the net of a cone cut out from $T^2/\mathbb{Z}_2$, whereas in the right panel, the cone cut out by $T^2/\mathbb{Z}_2$ 
    and a section of embedded quarter of $S^2$ with radius $r/\sqrt{3}$ are drawn.}
    \label{fig:Z2}
\end{figure}

First of all, we focus on the fixed point $(x,y)=(0,0)$ on $T^2/\mathbb{Z}_2$ orbifold. 
The blow-up of the orbifold singularity is carried out geometrically as follows.
Taking into account the $\mathbb{Z}_2$ identification $z\simeq -z$, a cone with radius $r/2$ is cut out, 
in which we embed a quarter of $S^2$ with radius $r/\sqrt{3}$ as illustrated in Fig.~\ref{fig:Z2}. 

Now, let us study the wavefunctions on the blow-up through the above geometrical procedure.
As discussed in Ref.~\cite{Conlon:2008qi}, the zero-mode wavefunction of scalar field on $\mathbb{CP}^1\simeq S^2$ with magnetic flux $M'$ is 
given by
\begin{align}
\phi^{j, M'}_{S^2} (z') = \frac{f^{j, M'}(z')}{\left(1+|z'|^2\right)^{M'/2}},
\end{align}
where $f^{j, M'}(z')$ is the holomorphic function with respect to the coordinate of $\mathbb{CP}^1$ $z'$ 
and $j$ denotes the multiplicities of zero-modes determined by the flux $M'$. 
Here, we use the same index label $j$ as in the $T^2/\mathbb{Z}_2$ case in order to smoothly connect the $S^2$ and 
$T^2/\mathbb{Z}_2$ wavefunctions in the blow-down limit $r\rightarrow 0$ as shown in the following analysis. 
The wavefunction of fermionic zero-mode on $\mathbb{CP}^1\simeq S^2$  is given by the same functional form, but the flux is shifted to 
$M'\rightarrow M'+1$ due to the curvature of $S^2$.
(That is, the form of wavefunctions is the same between the scalar and spinor fields, but the meaning of $M'$ 
is different by the spin connection. )
 In the following analysis, we focus on the scalar wavefunction. 
In the setup of Fig.~\ref{fig:Z2}, $z'$ corresponds to the $S^2$ coordinate as
\begin{align}
\left(\cos \frac{\theta}{2}\right)^{M'}=\frac{1}{\left(1+|z'|^2\right)^{M'/2}},
\end{align}
with $z'=\tan \frac{\theta}{2}e^{i\varphi}$. 
When we denote by $w$ the coordinate of a spherical surface, $w$ is written by $w=\frac{r}{2}\sqrt{3}z'=\frac{r}{2}e^{i\varphi}$ 
at $\theta =\theta_0$ as shown in Fig.~\ref{fig:Z2}. 
Furthermore, under the coordinate transformation $z\rightarrow w$, 
the derivative of holomorphic function $f^{j, M'}$ is transformed as
\begin{align}
\frac{d f^{j, M'}(z)}{d z}\biggl|_{z=r e^{i\varphi/2}} = \frac{d f^{j, M'}(w)}{d w}\biggl|_{w=\frac{r}{2} e^{i\varphi}} = 
\frac{1}{\frac{\sqrt{3}}{2}r} \frac{d f^{j, M'}(\frac{\sqrt{3}}{2}r z')}{d z'}\biggl|_{z'=\frac{1}{\sqrt{3}} e^{i\varphi}}. 
\label{eq:derivf}
\end{align}

Let us examine whether the $T^2/\mathbb{Z}_2$ wavefunction at $|z|=r$ is smoothly connected with 
the $S^2$ wavefunction at $\theta=\pi/3$ or not, namely
\begin{align}
c_1^{j, M}\phi_{T^2/\mathbb{Z}_2^+}^{j, M}\biggl|_{|z|=r} = d_1^{j, M'} \phi^{j, M'}_{S^2} (z')\biggl|_{\theta=\frac{\pi}{3}},
\end{align}
where we introduce the normalization factor $c_1^{j, M}$ and $d_1^{j, M'}$ and the bosonic wavefunction is now considered. 
As explicitly discussed later, the $\mathbb{Z}_2$-odd mode is not a zero-mode after the blow-up. 
To smoothly connect the $T^2/\mathbb{Z}_2$ wavefunction into the $S^2$ wavefunction in the blow-down limit $r\rightarrow 0$, 
we make the Ansatz that the holomorphic part of $T^2/\mathbb{Z}_2$ wavefunction (\ref{eq:T2Z2}) 
is unchanged after the blow-up. 

Under this Ansatz, we first investigate the non-holomorphic part of the $S^2$ and $T^2/\mathbb{Z}_2$ wavefunctions. 
We find that both the $T^2/\mathbb{Z}_2$ and $S^2$ wavefunctions are smoothly connected under the following two conditions,
\begin{align}
c_1^{j, M}e^{-\frac{\pi M}{2{\rm Im}\tau}r^2} &= d_1^{j, M'}\cos^{M'} \frac{\pi}{6},
\nonumber\\
\frac{d}{dr} c_1^{j, M}e^{-\frac{\pi M}{2{\rm Im}\tau}r^2} &= \frac{\sqrt{3}}{r} \frac{d}{d\theta} d_1^{j, M'}\cos^{M'} \frac{\theta}{2} \biggl|_{\theta = \frac{\pi}{3}},
\end{align}
which can be solved as
\begin{align}
\frac{c_1^{j, M}}{d_1^{j, M'}} = \left(\frac{\sqrt{3}}{2}\right)^{M'}e^{\frac{M'}{4}},\qquad 
\frac{M'}{4}=\frac{\pi r^2}{2{\rm Im}\tau}M.
\label{eq:ratio_N}
\end{align}
The latter flux condition is justified as follows. 
We remind that an amount of flux quanta in the $T^2/\mathbb{Z}_2$ region reduces to $\frac{\pi r^2/2}{{\rm Im}\tau}M$, because 
the area cut out from $T^2/\mathbb{Z}_2$ is $\pi r^2/2$ compared with the total area ${\rm Im}\tau$. 
By a similar reason, an amount of flux quanta in the $S^2$ region also reduces to $M'/4$ due to 
the fact that we embed a quarter of $S^2$.

Next, we discuss the holomorphic part of the wavefunctions. 
In the case of $j = 0, M/2$, we find the following holomorphic function $f^{j, M'}(z')$ in the bosonic wavefunction on $S^2$ 
is consistent with the coordinate transformation (\ref{eq:derivf}),
\begin{align}
f^{j, M'} (z') = g^{j, M}\left(\frac{\sqrt{3}}{2}r z'\right)
\end{align}
with 
\begin{align}
g^{j, M}(z') \equiv e^{\frac{\pi M}{2{\rm Im}\tau}z'^2}
\vartheta
\begin{bmatrix}
\frac{j}{M}\\
0
\end{bmatrix}
(Mz', M\tau) 
.
\end{align}
It is noted that $M'$ depends on $M$ through Eq.~(\ref{eq:ratio_N}).

On the other hand, in the case with $0 < j < M/2$, by using 
\begin{align}
\vartheta
\begin{bmatrix}
\frac{M-j}{M}\\
0
\end{bmatrix}
(Mz', M\tau) 
& = \sum_l e^{\pi i M\tau \left(\frac{M-j}{M} + l\right)^2} e^{2\pi i M z \left( \frac{M-j}{M} +l\right)} \nonumber\\
& = \sum_{l' = -l-1} e^{\pi i M\tau \left(\frac{j}{M} + l'\right)^2} e^{-2\pi i M z \left( \frac{j}{M} +l'\right)} \nonumber\\
& = \vartheta
\begin{bmatrix}
\frac{j}{M}\\
0
\end{bmatrix}
(Mz e^{i\pi}, M\tau) 
\end{align}
and taking into account the fact that the argument of $z'$ is multiplied by 2 under $z\rightarrow z'$, 
the holomorphic function $f$ is chosen as 
\begin{align}
f^{j, M'} (z') = \sqrt{2} g^{j, M}\left(\frac{\sqrt{3}}{2}r z'\right)
\end{align}
for $\mathbb{Z}_2$-even mode and $0$ for $\mathbb{Z}_2$-odd mode. 
Thus, only $\mathbb{Z}_2$-even mode is uplifted to $S^2$ which is consistent with the number of chiral zero-modes on 
both $T^2/\mathbb{Z}_2$ and $S^2$.

As a result, the wavefunction after the blow-up is described by 
\begin{align}
\phi^{j,M}_{\rm up} = 
\left\{
\begin{array}{c}
\phi^{j,M'}_{S^2} = \cfrac{d_1^{j,M'}}{\left(1+|z'|^2\right)^{\frac{M'}{2}}}g^{j, M}\left(\frac{r}{2}\sqrt{3} z'\right)\qquad (|z'|\leq \frac{1}{\sqrt{3}})
\\  
\phi^{j, M}_{T^2/\mathbb{Z}_2^+}= c_1^{j,M} e^{-\frac{\pi M |z|^2}{2{\rm Im}\tau}} g^{j, M}(z)\qquad (|z|\geq r)
\end{array}
\right.
,
\end{align}
for $j = 0, M/2$ and 
\begin{align}
\phi^{j,M}_{\rm up} = 
\left\{
\begin{array}{c}
\phi^{j,M'}_{S^2} = \cfrac{\sqrt{2}d_1^{j,M'}}{\left(1+|z'|^2\right)^{\frac{M'}{2}}}g^{j, M}\left(\frac{r}{2}\sqrt{3} z'\right)\qquad (|z'|\leq \frac{1}{\sqrt{3}})
\\  
\phi^{j, M}_{T^2/\mathbb{Z}_2^+}= c_1^{j,M} e^{-\frac{\pi M |z|^2}{2{\rm Im}\tau}} \cfrac{g^{j, M}(z) + g^{M-j, M}(z)}{\sqrt{2}}\qquad (|z|\geq r)
\end{array}
\right.
,
\end{align}
for $0 < j < M/2$. Although the explicit values of normalization factors $c_1^{j, M}$ and $d_1^{j, M'}$ are not determined yet, 
the ratio of the normalization factors is constrained as in Eq.~(\ref{eq:ratio_N}). 
The fermionic wavefunction after the blow-up is obtained by shifting the flux on $S^2$ as $M''=M'+1$, leading 
to $\cos^{M''-1}\frac{\theta}{2} = \cos^{M'}\frac{\theta}{2}$.

Note that the deficit angles of other three fixed points are all the same with the above case, namely a quarter of $S^2$ with radius $r/\sqrt{3}$ 
is embedded into each of the other fixed points. Totally, all the $S^2$ region is pasted together, meaning that the blow-up region is homeomorphic to $S^2$.

\subsubsection{$T^2/\mathbb{Z}_N$}
\begin{figure}[htbp]
\centering
  \begin{minipage}{0.3\columnwidth}
  \centering
     \includegraphics[scale=0.6]{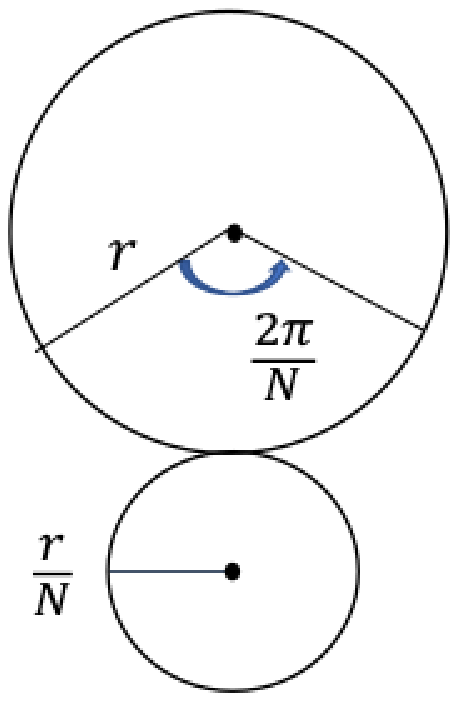}
   \end{minipage}
  \begin{minipage}{0.3\columnwidth}
   \centering
    \includegraphics[scale=0.6]{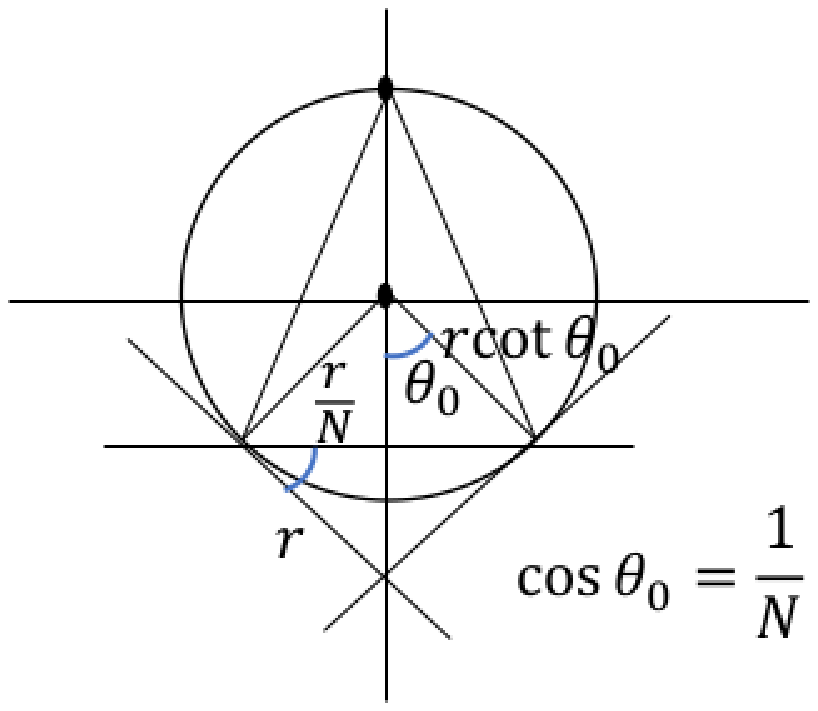}
  \end{minipage}
    \caption{The left panel shows the net of a cone cut out around the $\mathbb{C}/\mathbb{Z}_N$ singularity from $T^2/\mathbb{Z}_N$, whereas in the right panel, the cone cut out by $T^2/\mathbb{Z}_N$ 
    and a section of embedded subpart of $S^2$ with radius $\frac{r}{2\sqrt{2}}$ are drawn.}
    \label{fig:ZN}
\end{figure}

In this section, we generalize the $T^2/\mathbb{Z}_2$ to the $T^2/\mathbb{Z}_N$ system.
In particular, we focus on the $\mathbb{C}/\mathbb{Z}_N$ singularity on $T^2/\mathbb{Z}_N$.
Similar to the  $T^2/\mathbb{Z}_2$ orbifold, we cut out the region 
with radius $r$ around $\mathbb{C}/\mathbb{Z}_N$ singularity, where $\frac{N-1}{2N}$ of $S^2$ with radius $\frac{r}{N}$ is embedded as drawn in Fig.~\ref{fig:ZN}. 

To smoothly connect the $T^2/\mathbb{Z}_N$ and $S^2$ wavefunctions, we introduce the normalization factors $c_{N-1}^{j, M}$ and $d_{N-1}^{j, M'}$,
\begin{align}
c_{N-1}^{j, M}\phi_{T^2/\mathbb{Z}_N}^{j, M}\biggl|_{|z|=r} = d_{N-1}^{j, M'} \phi^{j,M'}_{S^2}(z')\biggl|_{\theta=\theta_0},
\end{align}
where we consider the bosonic wavefunction. 
Under our Ansatz that the holomorphic part of $T^2/\mathbb{Z}_N$ wavefunction is unchanged after the blow-up, 
both the $T^2/\mathbb{Z}_N$ and $S^2$ wavefunctions are smoothly connected under
\begin{align}
c_{N-1}^{j, M}e^{-\frac{\pi M}{2{\rm Im}\tau}r^2} &= d_{N-1}^{j, M'}\cos^{M'} \frac{\theta_0}{2},
\nonumber\\
\frac{d}{dr} c_{N-1}^{j, M}e^{-\frac{\pi M}{2{\rm Im}\tau}r^2}&=\frac{\tan \theta_0}{r} \frac{d}{d\theta} d_{N-1}^{j, M'}\cos^{M'} \frac{\theta}{2}\biggl|_{\theta=\theta_0}, 
\end{align}
with $\cos \theta_0 = \frac{1}{N}$, from which the normalization factors $c_{N-1}^{j, M}$, $d_{N-1}^{j, M'}$ and flux quanta are constrained as
\begin{align}
\frac{d_{N-1}^{j, M'}}{c_{N-1}^{j, M}} = \left(\frac{N+1}{2N}\right)^{M'/2}e^{\frac{N-1}{4}M'},\qquad 
\frac{N-1}{2N}M'=\frac{\pi r^2}{N{\rm Im}\tau}M.
\label{eq:ZNrelation}
\end{align}

We recall that an amount of flux quanta in $T^2/\mathbb{Z}_N$ region reduces to $\frac{\pi r^2}{N{\rm Im}\tau}M$, because 
the area cut out from $T^2/\mathbb{Z}_N$ is $\pi r^2/N$ compared with the total area ${\rm Im}\tau$. 
By a similar reason, an amount of flux quanta in the $S^2$ region also reduces to $(N-1)M'/(2N)$ due to 
the fact that we embed a proper part of $S^2$. 

The holomorphic part of the wavefunctions on $S^2$ should be consistent with the coordinate 
transformation,
\begin{align}
\frac{d f^{j, M'}(z)}{d z}\biggl|_{z=r e^{i\varphi/2}} = \frac{d f^{j, M'}(w)}{d w}\biggl|_{w=\frac{r}{2} e^{i\varphi}} = 
\frac{1}{\frac{r}{N}\sqrt{\frac{N+1}{N-1}}} \frac{d f^{j, M'}\left(\frac{r}{N}\sqrt{\frac{N+1}{N-1}} z'\right)}{d z'}\Biggl|_{z'=\sqrt{\frac{N+1}{N-1}} e^{i\varphi}},
\label{eq:derivfZN}
\end{align}
with $w=\frac{r}{N}\sqrt{\frac{N+1}{N-1}}z'$ being the coordinate of a spherical surface, 
and we then find that the holomorphic function $f^{j, M'}(z')$ in the bosonic wavefunction on $S^2$ is chosen as 
\begin{align}
f^{j, M'} (z') = \sqrt{N} g^{j, M}\left(\frac{r}{N}\sqrt{\frac{N+1}{N-1}}z'\right),
\end{align}
where it is noted that $M'$ depends on $M$ through Eq.~(\ref{eq:ZNrelation}). 
Following the same procedure in Sec.~\ref{subsubsec:T2Z2}, it turns out that only $\mathbb{Z}_N$-invariant mode 
functions ($m=0$ in Eq.~(\ref{eq:T2ZNfunc})) can be uplifted to $S^2$, meaning that other mode functions ($m\neq 0$) 
vanish. 

As a result, the wavefunction after the blow-up is obtained as
\begin{align}
\phi^{j,M}_{\rm up} = 
\left\{
\begin{array}{c} 
\phi^{j,M'}_{S^2} = \cfrac{\sqrt{N}d_{N-1}^{j, M'}}{\left(1+|z'|^2\right)^{\frac{M'}{2}}}g^{j, M}\left(\frac{r}{N}\sqrt{\frac{N+1}{N-1}}z'\right) \qquad (|z'|\leq \sqrt{\frac{N-1}{N+1}})
\\  
\phi^{j, M}_{T^2/\mathbb{Z}_N}= c_{N-1}^{j, M} e^{-\frac{\pi M |z|^2}{2{\rm Im}\tau}} \cfrac{1}{\sqrt{N}} \sum_{k=0}^{N-1} 
g^{j, M}(\rho^k z) \qquad (|z|\geq r)
\end{array}
\right.
.
\label{eq:T2Znwavefcn}
\end{align}

When we consider blow-ups of all the fixed points, 
the pasted region is homeomorphic to $S^2$.

\subsection{Normalization}
\label{subsec:2_3}

In this section, we determine the normalization factors.
We study the wavefunctions on the blow-up of $T^2/\mathbb{Z}_2$.
In particular, we consider the blow-up, where the singularity at only the origin $(x,y)=(0,0)$ is 
resolved. 
That is, we calculate the norm of the wavefunctions: 
\begin{align}
f_{jk} &\equiv 
\int_{\left|z\right| \geq r} dzd\bar{z} \ \phi^{j,M}_{T^2/\mathbb{Z}_2^+}\left(\phi^{k,M}_{T^2/\mathbb{Z}_2^+}\right)^{\ast} + \int_{\left|z'\right| \leq \frac{1}{\sqrt{3}}} dz'd\bar{z'}  \ \phi^{j,M'}_{S^{2}}\left(\phi^{k,M'}_{S^{2}}\right)^{\ast}
\nonumber\\
&=\left|\frac{c_1^{j, M}}{{\cal N}_M}\right|^2\delta_{jk} - \int_{\left|z\right| \leq r} dzd\bar{z} \ \phi^{j,M}_{T^2/\mathbb{Z}_2^+}\left(\phi^{k,M}_{T^2/\mathbb{Z}_2^+}\right)^{\ast} 
+ \int_{\left|z'\right| \leq \frac{1}{\sqrt{3}}} dz'd\bar{z'}  \ \phi^{j,M'}_{S^{2}}\left(\phi^{k,M'}_{S^{2}}\right)^{\ast}, 
\label{eq:normalization} 
\end{align}
where ${\cal N}_M=(2M{\rm Im}(\tau)/{\cal A}^2)^{1/4}$ is the normalization factor on $T^2/\mathbb{Z}_2$. 
Here and in what follows, we show the bosonic wavefunctions, but it is applicable to the fermionic wavefunctions 
by shifting the flux $M\rightarrow M + 1$. 

Details of computations are shown in Appendix~\ref{app}.
As a result, the normalization factor reduces to be
\begin{align}
f_{jk} &\simeq 
\left\{
\begin{array}{c}
\frac{c_1^{j, M} (c_1^{k, M})^\ast}{|{\cal N}_M|^2}  \left(\delta_{jk}+\frac{1}{2\pi}\left(\frac{\pi r^2}{2}\left[\phi^{j,M}_{T^2/\mathbb{Z}_2^-}\right]' \left(0\right) \right) 
\left(\frac{\pi r^2}{2}\left[\phi^{k,M}_{T^2/\mathbb{Z}_2^-}\right]' \left(0\right)\right)^{\ast}\right)\qquad (0 < j < \frac{M}{2})
\\
\frac{c_1^{j, M} (c_1^{k, M})^\ast}{|{\cal N}_M|^2} \delta_{jk} \qquad ( j = 0, \frac{M}{2}) 
\end{array}
\right.
,
\label{eq:fjk}
\end{align}
where $\phi^{j,M}_{T^2/\mathbb{Z}_2^-}$ is the $\mathbb{Z}_2$-odd wavefunction on $T^2/\mathbb{Z}_2$ and we note that 
$\left[\phi^{j,M}_{T^2/\mathbb{Z}_2^+}\right]' \left(0\right) \equiv \frac{d\phi^{j,M}_{T^2/\mathbb{Z}_2^+}}{dz} \Biggl|_{z=0} = 0$. 
It is remarkable that the wavefunctions with $j=0, \frac{M}{2}$ still correspond to the orthogonal basis 
even on the blow-up of the $T^2/\mathbb{Z}_2$ orbifold.
Thus, we set $\left|c_1^{j, M}\right| \simeq \left|{\cal N}_M\right|$ for $j=0, \frac{M}{2}$.
In particular, the model with $M=2$ has only these two modes, and nothing changes even after the blow-up.

On the other hand, for $0 < j < \frac{M}{2}$, $\mathbb{Z}_2$-odd effects affect the $\mathbb{Z}_2$-even part at the ${\cal O}(r^4)$ level. 
When $r^4 \ll 1$, the effect of $f_{jk} \ \left(j \neq k\right)$ on normalization is much smaller 
than $f_{jj} \ \left(j=k\right)$.
Thus, we normalize the wavefunctions such that  they satisfy $f_{jj}=1$. 
Then, the normalization factor is given by
\begin{eqnarray}
\left|c_1^{j, M}\right| &=& \left|{\cal N}_M\right| \left(1 + \frac{1}{\pi} \left(\frac{\pi r^2}{2}\right)^2 \left|\left[\phi^{j,M}_{T^2/\mathbb{Z}_2^-}\right]' \left(0\right)\right|^2 \right)^{-\frac{1}{2}} \notag \\
&\simeq& \left|{\cal N}_M\right| \left(1 - \frac{1}{2\pi} \left(\frac{\pi r^2}{2}\right)^2 \left|\left[\phi^{j,M}_{T^2/\mathbb{Z}_2^-}\right]' \left(0\right)\right|^2 \right). 
\label{eq:Normalizationapp}
\end{eqnarray}

In addition, this $f_{jk}$ is a Hermitian matrix and unitary at ${\cal O}(r^4)$ order which gives 
the wavefunction $\tilde{\phi}^{j'}$ in the orthonormal basis. 
When we expand the wavefunction around the origin $z=z'=0$, like $\phi^j(z) \simeq \phi^j(0) +(\phi^j)'(0)z$, 
and rotate it at ${\cal O}(r^4)$, $\phi^j(0)$ can be rotated at the order of $r^4$, but $(\phi^j)'(0)$ is rotated at ${\cal O}(r^5)$ order due to 
the fact that $z$ linearly depends on $r$. Thus, it is enough to consider $\phi^j(0)$ at the ${\cal O}(r^4)$ level, in other words 
the wavefunctions are orthogonal to each other at $z, z' \neq 0$, namely $\tilde{\phi}^{j'}=\phi^j$. 
From the unitary matrix $f$, $\phi^j(0)$ can be expanded as
\begin{align}
\phi^{j}\left(0\right) = \tilde{\phi}^{j'}\left(0\right)+\frac{1}{2\pi}\left(\frac{\pi r^2}{2}\right)^2 \sum_{l' \neq j'}\left(\left[\tilde{\phi}^{j'}_{T^2/\mathbb{Z}_2^-}\right]'\left(0\right)\right)\left(\left[\tilde{\phi}^{l'}_{T^2/\mathbb{Z}_2^-}\right]'\left(0\right)\right)^{\ast}\left(\tilde{\phi}^{l'}\left(0\right)\right),
\end{align}
where the second term is relevant for $M\geq 5$. 
Indeed, only for these values of $M$, the $\mathbb{Z}_2$-odd pair gives $\mathbb{Z}_2$-even term.

Similarly, we can compute the wavefunction normalization on the blow-up of $T^2/\mathbb{Z}_2$, where 
some and all of fixed points are resolved.
Also, we can calculate the wavefunction normalization of the blow-ups of $T^2/\mathbb{Z}_N$ orbifolds.

\section{Yukawa couplings and modular symmetry}
\label{sec:3}
In this section, we study Yukawa couplings 
on resolutions of $T^2/\mathbb{Z}_2$ 
by calculating the overlap integral of the zero-mode wavefunctions. 
We also discuss the modular symmetry.

\subsection{Yukawa couplings on resolutions of $T^2/\mathbb{Z}_2$}
\label{subsec:3_1}
In the following analysis, we do not distinguish the bosonic and fermionic wavefunctions, since their 
functional forms are the same. 
By using the identity of theta function:
\begin{align}
\vartheta
\begin{bmatrix}
\frac{r}{N_1} \\ 0
\end{bmatrix}
\left(z_1,\tau N_1 \right) \vartheta
\begin{bmatrix}
\frac{s}{N_2} \\ 0
\end{bmatrix}
\left(\mathbb{Z}_2,\tau N_2 \right)
&=  \sum_{m \in \bf{Z}_{N_1+N_2}} \vartheta
\begin{bmatrix}
\frac{r+s+N_1m}{N_1+N_2} \\ 0
\end{bmatrix}
\left(z_1+\mathbb{Z}_2,\tau \left(N_1+N_2 \right) \right) \\
&\times \vartheta
\begin{bmatrix}
\frac{N_2r-N_1s+N_1N_2m}{N_1N_2\left(N_1+N_2\right)} \\ 0
\end{bmatrix}
\left(z_1N_2-\mathbb{Z}_2N_1,\tau N_1N_2 \left(N_1+N_2\right) \right) \notag 
,
\label{eq:thetaformula}
\end{align}
we obtain the zero-mode product expansion for $i, j=0, \frac{M}{2}$,
\begin{gather}
\phi^{i,I_{ab}}_{T^2/\mathbb{Z}_2^+}\phi^{j,I_{ca}}_{T^2/\mathbb{Z}_2^+} = \left|\frac{c^{i,I_{ab}}c^{j,I_{ca}}}{c^{i+j+I_{ab}m,I_{cb}}}\right| \sum_{m \in \bf{Z}_{I_{bc}}} \phi^{i+j+I_{ab}m,I_{cb}}_{T^2/\mathbb{Z}_2^+}
\vartheta
\begin{bmatrix}
\frac{I_{ca}i-I_{ab}j+I_{ab}I_{ca}m}{-I_{ab}I_{ca}I_{bc}} \\ 0
\end{bmatrix}
\left(0,\tau \left|I_{ab}I_{bc}I_{ca}\right|\right),
\nonumber\\
\phi^{i,I_{ab}}_{S^{2}}\phi^{j,I_{ca}}_{S^{2}} = \left|\frac{d^{i,I_{ab}}d^{j,I_{ca}}}{d^{i+j+I_{ab}m,I_{cb}}}\right| \sum_{m \in \bf{Z}_{I_{bc}}} \phi^{i+j+I_{ab}m,I_{cb}}_{S^{2}}
\vartheta
\begin{bmatrix}
\frac{I_{ca}i-I_{ab}j+I_{ab}I_{ca}m}{-I_{ab}I_{ca}I_{bc}} \\ 0
\end{bmatrix}
\left(0,\tau \left|I_{ab}I_{bc}I_{ca}\right|\right), 
\label{eq:productj=0} 
\end{gather}
where we denote magnetic fluxes by $I_{ab}, I_{bc}$, and $I_{ca}$.
By use of these identities, Yukawa couplings are derived as\footnote{See for details of computations, Appendix~\ref{app-2}.}
\begin{eqnarray}
Y_{ijk} 
&\simeq& \left|\frac{N_{I_{ab}}N_{I_{ca}}}{N_{I_{bc}}}\right|
\vartheta
\begin{bmatrix}
-\frac{1}{I_{ab}}\left(\frac{j}{I_{ca}}+\frac{k}{I_{bc}}\right) \\ 0
\end{bmatrix}
\left(0,\tau \left|I_{ab}I_{bc}I_{ca}\right|\right). 
\label{eq:Yukawa_main}
\end{eqnarray}
The obtained Yukawa couplings are unchanged after the blow-up, but we note that 
$\left|\frac{\left(\frac{d^{j, I_{ab}}}{c^{j, I_{ab}}}\right)\left(\frac{d^{j, I_{ca}}}{c^{j,I_{ca}}}\right)}{\left(\frac{d_{j, I_{cb}}}{c^{j, I_{cb}}}\right)}\right|=1$ using $I_{ab}+I_{ca}=I_{cb}=-I_{bc}$ and $k=i+j \ {\rm mod} \ I_{ab}$. 
Indeed, when $i,j=0,\frac{M}{2}$, the normalization constant is also the same with the toroidal orbifold one as in Eq.~(\ref{eq:fjk}).

By contrast, when $0 <j < \frac{M}{2}$, Eq.~(\ref{eq:productj=0}) is changed to
\begin{align}
\phi^{i,I_{ab}}_{T^2/\mathbb{Z}_2^{+}}\phi^{j,I_{ca}}_{T^2/\mathbb{Z}_2^{+}} &= 
\frac{1}{\sqrt{2}} \sum_{m \in \bf{Z}_{I_{bc}}} \left|\frac{c^{i,I_{ab}}c^{j,I_{ca}}}{c^{i+j+I_{ab}m,I_{cb}}}\right| \phi^{i+j+I_{ab}m,I_{cb}}_{T^2/\mathbb{Z}_2^{+}}
\vartheta
\begin{bmatrix}
\frac{I_{ca}i-I_{ab}j+I_{ab}I_{ca}m}{-I_{ab}I_{ca}I_{bc}} \\ 0
\end{bmatrix}
\left(0,\tau \left|I_{ab}I_{bc}I_{ca}\right|\right) 
\nonumber \\
&+ \left|\frac{c^{i,I_{ab}}c^{j,I_{ca}}}{c^{I_{ab}-i+j+I_{ab}m,I_{cb}}}\right|
\phi^{I_{ab}-i+j+I_{ab}m,I_{cb}}_{T^2/\mathbb{Z}_2^{+}}
\vartheta
\begin{bmatrix}
\frac{I_{ca}\left(I_{ab}-i\right)-I_{ab}j+I_{ab}I_{ca}m}{-I_{ab}I_{ca}I_{bc}} \\ 0
\end{bmatrix}
\left(0,\tau \left|I_{ab}I_{bc}I_{ca}\right|\right)
.
\end{align}
The identity for $\phi^{i,I_{ab}}_{S^{2}}$ changes by the factor $\sqrt{2}$ from Eq.~(\ref{eq:productj=0}).
These zero-mode product expansion is realized even if we include the ${\cal O}(r^4)$ correction.

By use of these zero-mode product expansions, the Yukawa couplings are calculated as
\begin{align}
Y_{ijk} 
=& \frac{\left|c^{i,I_{ab}}c^{j,I_{ca}}c^{k,I_{cb}}\right|}{\sqrt{2} \left|N_{I_{cb}}\right|^2}
\vartheta
\begin{bmatrix}
-\frac{1}{I_{ab}}\left(\frac{j}{I_{ca}}+\frac{k}{I_{bc}}\right) \\ 0
\end{bmatrix}
\left(0,\tau \left|I_{ab}I_{bc}I_{ca}\right|\right) + \frac{1}{4\pi} \left(\frac{\pi r^2}{2}\right)^2 \times\notag \\
& \left( 2\sqrt{2} \left[\phi^{i,I_{ab}}_{T^2}\left(0\right)\phi^{j,I_{ca}}_{T^2}\left(0\right)\left(\phi^{k,I_{cb}}_{T^2}\left(0\right)\right)^{\ast}\right]'' - \left[\phi^{i,I_{ab}}_{T^2/\mathbb{Z}_2^{+}}\left(0\right)\phi^{j,I_{ca}}_{T^2/\mathbb{Z}_2^{+}}\left(0\right)\left(\phi^{k,I_{cb}}_{T^2/\mathbb{Z}_2^{+}}\left(0\right)\right)^{\ast}\right]''\right)
.
\label{eq:Yukawa_main2}
\end{align}
The first term is coming from $k=i+j \ {\rm mod} \ I_{ab}$ or $k=I_{ab}-i+j \ {\rm mod} \ I_{ab}$. 
When $k=0, \frac{I_{cb}}{2}$, the right-handed side becomes $\sqrt{2}$ times Eq.~(\ref{eq:Yukawa_main2}) and the above 
Yukawa couplings are the same with Eq.~(\ref{eq:Yukawa_main}) when $i,j =0, \frac{I}{2}$. 
Thus, Yukawa couplings receive ${\cal O}(r^4)$ corrections except for $i,j =0, \frac{I}{2}$, 
where the ${\cal O}(r^4)$ corrections are mostly originating from the wavefunction at the origin.

Similarly, we can compute higher-order couplings by using zero-mode product expansions \cite{Abe:2009dr,Honda:2018sjy}.

\subsection{Modular symmetry}
\label{subsec:3_2}
In this section, we discuss the  modular symmetry, SL$(2, \mathbb{Z})$, which is generated by 
two elements, 
\begin{align}
S &: \tau \rightarrow -\frac{1}{\tau},
\nonumber\\
T &: \tau \rightarrow \tau +1.
\end{align}

As shown in Eq.~(\ref{eq:T2Znwavefcn}), the unnormalized wavefunction on the resolutions of $T^2/\mathbb{Z}_N$ orbifolds is written 
by the linear combination of that on torus $T^2$. 
Modular transformations of wavefunctions on $T^2$ with magnetic fluxes were shown 
in Refs.~\cite{Cremades:2004wa,Kobayashi:2017dyu,Kobayashi:2018rad,Kobayashi:2018bff}.

Let us first examine the $S$-transformation:
\begin{align}
\tau \rightarrow -\frac{1}{\tau}.
\end{align}
Around each fixed point $|z'|\leq \sqrt{\frac{N-1}{N+1}}$, the $S^2$ wavefunction transforms as
\begin{align}
 \phi^{j,M'}_{S^2} \left(\tau, z'\right) &= \frac{\sqrt{N}d_{N-1}^{j,M'}}{(1+|z'|^2)^{M'/2}}e^{\frac{(N+1)\pi Mr^2}{2N^2(N-1){\rm Im}\tau}(z')^2} 
 \vartheta
\begin{bmatrix}
\frac{j}{M}\\
0
\end{bmatrix}
\left(\frac{r}{N}\sqrt{\frac{N+1}{N-1}}Mz', M\tau\right) 
\nonumber\\
&\rightarrow 
\frac{\sqrt{N}d_{N-1}^{j,M'}}{(1+|z'|^2)^{M'/2}}e^{\frac{(N+1)\pi M r^2}{2N^2(N-1){\rm Im}\tau}
\left(\frac{|\tau| z'}{\tau}\right)^2} 
 \vartheta
\begin{bmatrix}
\frac{j}{M}\\
0
\end{bmatrix}
\left(\frac{r}{|\tau|N}\sqrt{\frac{N+1}{N-1}}M\frac{|\tau|z'}{\tau}, M\tau\right) 
\nonumber\\
&= (-i \tau)^{1/2} \frac{1}{\sqrt{M}} \frac{\sqrt{N}d_{N-1}^{j,M'}}{(1+|z'|^2)^{M'/2}}e^{\frac{(N+1)\pi M r^2}{2N^2(N-1){\rm Im}\tau}(z')^2} 
 \vartheta
\begin{bmatrix}
0\\
\frac{j}{M}
\end{bmatrix}
\left(-\frac{r}{N}\sqrt{\frac{N+1}{N-1}}z', \frac{\tau}{M}\right) 
\nonumber\\
&\equiv (-i \tau)^{1/2} \chi^{j,M'}_{S^2} \left(\tau, -z'\right)
\nonumber\\
&= (-i \tau)^{1/2}  \frac{1}{\sqrt{M'}} \sum_k e^{2\pi i \frac{jk}{M'}} \phi^{k,M'}_{S^2} \left(\tau, -z'\right),
\end{align}
whereas outside the fixed point $|z|\geq r$, we obtain
\begin{align}
 \phi^{j,M}_{T^2} \left(\tau, z\right) &=c_{N-1}^{j,M} e^{i\pi M z\frac{{\rm Im}z}{{\rm Im}\tau} }
 \vartheta
\begin{bmatrix}
\frac{j}{M}\\
0
\end{bmatrix}
\left(Mz, M\tau\right) 
\nonumber\\
 &\rightarrow c_{N-1}^{j,M'} e^{i\pi M \frac{z}{\tau} \frac{{\rm Im}\bar{\tau}z}{{\rm Im}\tau} }
 \vartheta
\begin{bmatrix}
\frac{j}{M} \\
0
\end{bmatrix}
\left(M\frac{z}{\tau}, -\frac{M}{\tau}\right) 
\nonumber\\
 &=(-i \tau)^{1/2} \frac{1}{\sqrt{M}} c_{N-1}^{j,M'} e^{i\pi M z \frac{{\rm Im}z}{{\rm Im}\tau} } 
 \vartheta
\begin{bmatrix}
0\\
\frac{j}{M}
\end{bmatrix}
\left(-z, \frac{\tau}{M}\right)
\nonumber\\
 &\equiv (-i \tau)^{1/2} \chi^{j,M}_{T^2} \left(\tau, -z\right) 
 \nonumber\\
 &=(-i \tau)^{1/2} \frac{1}{\sqrt{M}} \sum_k e^{2\pi i \frac{jk}{M}} \phi^{k,M}_{T^2} \left(\tau, -z'\right).
\end{align}
Here, we used
\begin{equation}
\vartheta
\begin{bmatrix}
0 \\ a
\end{bmatrix}
\left(\frac{\nu}{\kappa}, -\frac{1}{\kappa} \right) =
\left(-i\kappa \right)^{\frac{1}{2}} e^{i\pi \frac{\nu^2}{\kappa}} \vartheta
\begin{bmatrix}
a \\ 0
\end{bmatrix}
\left(\nu, \kappa \right) \label{thetaformula}
\end{equation}
and employed
\begin{align}
\chi^{j,M} &=\frac{1}{\sqrt{M}} \sum_k e^{2\pi i \frac{jk}{M}} \phi^{k,M} .
\end{align}
In addition, since the $T$ generator transforms only the theta function part, 
 the $S^2$ wavefunction and the $T^2/\mathbb{Z}_2$ wavefunction outside the fixed point $|z|\geq r$ 
 transform in the same way under the $T$-transformation.
Thus, unnormalized wavefunctions transform under both $S$- and $T$-transformations in the same way 
as those without the blow-up.  
However, in the orthonormal basis of wavefunctions as demonstrated in $T^2/\mathbb{Z}_2$ case (Sec.~\ref{subsec:2_3}), 
normalization factor receives ${\cal O}(r^4)$ corrections which are changed under the modular transformations. 
Because of these corrections, modular transformation behaviors of wavefunctions on the blow-up 
of $T^2/\mathbb{Z}_2$ are different from those on the orbifold $T^2/\mathbb{Z}_2$ by  ${\cal O}(r^4)$.

\section{Conclusion}
\label{sec:con}
We have proposed the zero-mode wavefunctions on the resolutions of $T^2/\mathbb{Z}_N$ orbifolds with constant 
magnetic fluxes, where the orbifold fixed points are replaced by a part of sphere. 
We find that the $\mathbb{Z}_N$-invariant zero-mode wavefunctions on $T^2/\mathbb{Z}_N$ orbifolds are smoothly connected with wavefunctions on $S^2$. 
Since obtained zero-mode wavefunctions receive blow-up effects at the order of $r^4$, where $r$ is the blow-up radius, 
the modular transformation of wavefunctions and the  Yukawa couplings among chiral zero-modes are different from 
the toroidal orbifold results. 
It is interesting to extent our results to more general higher-dimensional toroidal orbifolds such as $T^4/\mathbb{Z}_N$ and 
$T^6/\mathbb{Z}_N$.

\section*{Acknowledgments}
T.~K. was supported in part by MEXT KAKENHI Grant Number JP19H04605.
  H.~O. was supported in part by Grant-in-Aid for JSPS Research Fellow 
  from Japan Society for the Promotion of Science.

\appendix 

\section{Normalization of wavefunctions and Yukawa couplings}
\label{app}

\subsection{Normalization}
\label{app-1}

Here, we calculate $f_{jk}$ in Eq.~(\ref{eq:normalization}) for $0 < j < \frac{M}{2}$. 
Under the blow-down regime $\frac{\pi r^2}{{\rm Im}\tau} \ll 1$, normalization factor is estimated as
\begin{align}
&\int_{\left|z\right| \leq r} dzd\bar{z} \ \phi^{j,M}_{T^2/\mathbb{Z}_2^{+}}\left(\phi^{k,M}_{T^2/\mathbb{Z}_2^{+}}\right)^{\ast} \notag \\
&\simeq \frac{c_1^{j, M} (c_1^{k, M})^\ast}{2} \sum_{l,m} e^{\pi iMRe\tau\left[\left(\frac{j}{M}+l \right)^2-\left(\frac{k}{M}+m \right)^2\right]} e^{-\pi MIm\tau \left[\left(\frac{j}{M}+l\right)^2+\left(\frac{k}{M}+m\right)^2 \right]} \int_0^r d\left|z\right| \left|z\right| \int_{-\frac{\pi}{2}}^{\frac{\pi}{2}} d\left(\frac{\varphi}{2}\right) \notag \\
&\times \bigl[ e^{2\pi iM\left|z\right|\cos \frac{\varphi}{2}\left[\left(\frac{j}{M}+l \right)-\left(\frac{k}{M}+m \right)\right]} + e^{-2\pi iM\left|z\right|\cos \frac{\varphi}{2}\left[\left(\frac{j}{M}+l \right)-\left(\frac{k}{M}+m \right)\right]} \notag \\
&+ e^{2\pi iM\left|z\right|\cos \frac{\varphi}{2}\left[\left(\frac{j}{M}+l \right)+\left(\frac{k}{M}+m \right)\right]} + e^{-2\pi iM\left|z\right|\cos \frac{\varphi}{2}\left[\left(\frac{j}{M}+l \right)+\left(\frac{k}{M}+m \right)\right]} \bigl] \notag \\
&\simeq c_1^{j, M} (c_1^{k, M})^\ast \sum_{l,m} e^{\pi iM\left[\tau \left(\frac{j}{M}+l \right)^2-\bar{\tau} \left(\frac{k}{M}+m \right)^2\right]} \int_0^r d\left|z\right| \left|z\right| \left(2\pi-2\pi^3 \left|z\right|^2 \left[\left(j+lM \right)^2+\left(k+mM \right)^2\right]\right) \notag \\
&= c_1^{j, M} (c_1^{k, M})^\ast \pi r^2 \sum_{l,m} e^{\pi iM\left[\tau \left(\frac{j}{M}+l \right)^2-\bar{\tau} \left(\frac{k}{M}+m \right)^2\right]} \left(1-\frac{\left(\pi r\right)^2}{2}  \left[\left(j+lM \right)^2+\left(k+mM \right)^2\right] \right) 
\label{eq:Z2evennorm}
\end{align}
and 
\begin{eqnarray}
&&\int_{\left|z'\right| \leq \frac{1}{\sqrt{3}}} dz'd\bar{z'} \ \phi^{j,M'}_{S^{2}}\left(\phi^{k,M'}_{S^{2}}\right)^{\ast} \notag \\
&\simeq& d_1^{j, M'} (d_1^{k, M'})^\ast \sum_{l,m} 2e^{\pi iM\left[\tau \left(\frac{j}{M}+l \right)^2-\bar{\tau} \left(\frac{k}{M}+m \right)^2\right]} \int_0^{\frac{1}{\sqrt{3}}} \left(r\frac{\sqrt{3}}{2}d\left|z'\right|\right)r\frac{\sqrt{3}}{2}\left|z'\right| \frac{e^{\frac{3}{4}\frac{\pi r^2}{Im\tau}M\left|z'\right|^2}}{\left(1+\left|z'\right|^2\right)^{\frac{2\pi r^2}{Im\tau}M}} \notag \\
&\times& \int_{-\pi}^{\pi} d\varphi \ e^{2\pi iMr\frac{\sqrt{3}}{2}\left|z'\right|\cos \varphi \left[\left(\frac{j}{M}+l \right)-\left(\frac{k}{M}+m \right)\right]} \notag \\
&=& d_1^{j, M'} (d_1^{k, M'})^\ast \sum_{l,m} e^{\pi iM\left[\tau \left(\frac{j}{M}+l \right)^2-\bar{\tau} \left(\frac{k}{M}+m \right)^2\right]} \int_0^{\frac{1}{\sqrt{3}}} \left(r\frac{\sqrt{3}}{2}d\left|z'\right|\right)r\frac{\sqrt{3}}{2}\left|z'\right| \frac{e^{\frac{3}{4}\frac{\pi r^2}{Im\tau}M\left|z'\right|^2}}{\left(1+\left|z'\right|^2\right)^{\frac{2\pi r^2}{Im\tau}M}} \notag \\
&\times& 2\int_{0}^{\pi} d\varphi \ \cos \left(2\pi Mr\frac{\sqrt{3}}{2}\left|z'\right|\cos \varphi \left[\left(\frac{j}{M}+l \right)-\left(\frac{k}{M}+m \right)\right]\right) \notag \\
&\simeq & d_1^{j, M'} (d_1^{k, M'})^\ast \sum_{l,m} 2e^{\pi iM\left[\tau \left(\frac{j}{M}+l \right)^2-\bar{\tau} \left(\frac{k}{M}+m \right)^2\right]} \int_0^{\frac{1}{\sqrt{3}}} \left(r\frac{\sqrt{3}}{2}d\left|z'\right|\right)r\frac{\sqrt{3}}{2}\left|z'\right| \frac{e^{\frac{3}{4}\frac{\pi r^2}{Im\tau}M\left|z'\right|^2}}{\left(1+\left|z'\right|^2\right)^{\frac{2\pi r^2}{Im\tau}M}} \notag \\
&\times& \left(\left(2\pi\right)-\left(2\pi\right)\pi^2\left(r\frac{\sqrt{3}}{2}\left|z'\right|\right)^2\left[\left(j+lM \right)-\left(k+mM \right)\right]^2 \right) \notag \\
&\simeq& c_1^{j, M} (c_1^{k, M})^\ast \pi r^2 \sum_{l,m} e^{\pi iM\left[\tau \left(\frac{j}{M}+l \right)^2-\bar{\tau} \left(\frac{k}{M}+m \right)^2\right]} \left(1-\frac{\left(\pi r\right)^2}{2} \left[\left(j+lM \right)-\left(k+mM \right)\right]^2 \right).
\label{eq:S2norm}
\end{eqnarray}

In a similar way, in $j=0, \frac{M}{2}$ case, normalization factor becomes
\begin{eqnarray}
&&\int_{\left|z\right| \leq r} dzd\bar{z} \ \phi_{T^2/\mathbb{Z}_2^+}^{j,M} \left(\phi_{T^2/\mathbb{Z}_2^+}^{j,M}\right)^{\ast} \notag \\
&\simeq& c_1^{j, M} (c_1^{k, M})^\ast \sum_{l,m} e^{\pi iM \left[\tau\left(\frac{j}{M}+l\right)^2-\bar{\tau}\left(\frac{k}{M}+m\right)^2\right]}
\int_0^r d\left|z\right| \left|z\right| \int_{-\frac{\pi}{2}}^{\frac{\pi}{2}} d\left(\frac{\varphi}{2}\right) e^{2\pi iM \left|z\right| \cos \frac{\varphi}{2} \left[\left(\frac{j}{M}+l\right)-\left(\frac{k}{M}+m\right)\right]} \notag \\
&\simeq& c_1^{j, M} (c_1^{k, M})^\ast \sum_{l,m} e^{\pi iM \left[\tau\left(\frac{j}{M}+l\right)^2-\bar{\tau}\left(\frac{k}{M}+m\right)^2\right]}
\int_0^r d\left|z\right| \left|z\right| \left(\pi- \pi^3 \left|z\right|^2 \left[\left(j+lM\right)-\left(k+mM\right)\right]^2 \right) \notag \\
&=& c_1^{j, M} (c_1^{k, M})^\ast \frac{\pi r}{2} \sum_{l,m} e^{\pi iM \left[\tau\left(\frac{j}{M}+l\right)^2-\bar{\tau}\left(\frac{k}{M}+m\right)^2\right]}
\left(1-\frac{\left(\pi r\right)^2}{2} \left[\left(j+lM\right)-\left(k+mM\right)\right]^2 \right), 
\label{eq:evenonlynorm}
\end{eqnarray}
which is consistent with Eq.~(\ref{eq:Z2evennorm}) by taking into account $\frac{j}{M}+l \rightarrow -\left(\frac{j}{M}+l\right)$ 
and the multiplication of overall factor$\left(\frac{1}{\sqrt{2}}\right)^2$ in Eq.~(\ref{eq:Z2evennorm}). 
Furthermore, $\phi^{j,M'}_{S^{2}}$ part is determined by multiplying $\left(\frac{1}{\sqrt{2}}\right)^2$ with Eq.~(\ref{eq:S2norm}),
\begin{align}
&\int_{\left|z'\right| \leq \frac{1}{\sqrt{3}}} dz'd\bar{z'} \ \phi^{j,M'}_{S^{2}}\left(\phi^{k,M'}_{S^{2}}\right)^{\ast} \notag \\
&\simeq c_1^{j, M} (c_1^{k, M})^\ast \frac{\pi r^2}{2} \sum_{l,m} e^{\pi iM\left[\tau \left(\frac{j}{M}+l \right)^2-\bar{\tau} \left(\frac{k}{M}+m \right)^2\right]} \left(1-\frac{\left(\pi r\right)^2}{2} \left[\left(j+lM \right)-\left(k+mM \right)\right]^2 \right).
\label{eq:S2evenonlynorm}
\end{align} 
When $0<j<\frac{M}{2}$ and $j=0, \frac{M}{2}$ cases appear in the integral, we can calculate $T^2/\mathbb{Z}_2$ part by replacing 
$\frac{j}{M}+l \rightarrow -\left(\frac{j}{M}+l\right)$ for $j=0, \frac{M}{2}$ and the result is just $\sqrt{2}$ times Eq.~(\ref{eq:evenonlynorm}). 
In the $S^2$ part, the result is also $\sqrt{2}$ times Eq.~(\ref{eq:S2evenonlynorm}).

\subsection{Yukawa couplings}
\label{app-2}

By use of Eq.~(\ref{eq:productj=0}),  Yukawa couplings are derived as
\begin{align}
Y_{ijk} &= \int_{\left|z\right| \geq r} dzd\bar{z} \ \phi^{i,I_{ab}}_{T^2/\mathbb{Z}_2^+}\phi^{j,I_{ca}}_{T^2/\mathbb{Z}_2^+}\left(\phi^{k,I_{cb}}_{T^2/\mathbb{Z}_2^+}\right)^{\ast} + \int_{{\left|z'\right| \leq \frac{1}{\sqrt{3}}}} dz'd\bar{z'} \ \phi^{i,I_{ab}}_{S^{2}}\phi^{j,I_{ca}}_{S^{2}}\left(\phi^{k,I_{cb}}_{S^{2}}\right)^{\ast} \notag\\
&= \sum_{m \in \bf{Z}_{I_{bc}}} \vartheta
\begin{bmatrix}
\frac{I_{ca}i-I_{ab}j+I_{ab}I_{ca}m}{-I_{ab}I_{ca}I_{bc}} \\ 0
\end{bmatrix}
\left(0,\tau \left|I_{ab}I_{bc}I_{ca}\right|\right) \times \biggl[
\left|\frac{c^{i,I_{ab}}c^{j,I_{ca}}}{c^{k,I_{cb}}}\right| \int_{\left|z\right| \geq r} dzd\bar{z} \ \phi^{i+j+I_{ab}m,I_{cb}}_{T^2/\mathbb{Z}_2^+}\left(\phi^{k,I_{cb}}_{T^2/\mathbb{Z}_2^+}\right)^{\ast}
\notag\\
&+ \left|\frac{d^{i,I_{ab}}d^{j,I_{ca}}}{d^{k,I_{cb}}}\right| \int_{{\left|z'\right| \leq \frac{1}{\sqrt{3}}}} dz'd\bar{z'} \ \phi^{i+j+I_{ab}m,I_{cb}}_{S^{2}} \left(\phi^{k,I_{cb}}_{S^{2}}\right)^{\ast}\biggl] \notag\\
&= \left|\frac{c^{i,I_{ab}}c^{j,I_{ca}}}{c^{k,I_{cb}}}\right| \sum_{m \in \bf{Z}_{I_{bc}}} \vartheta
\begin{bmatrix}
\frac{I_{ca}i-I_{ab}j+I_{ab}I_{ca}m}{-I_{ab}I_{ca}I_{bc}} \\ 0
\end{bmatrix}
\left(0,\tau \left|I_{ab}I_{bc}I_{ca}\right|\right) \notag\\
&\times \left[
\int_{\left|z\right| \geq r} dzd\bar{z} \ \phi^{i+j+I_{ab}m,I_{cb}}_{T^2/\mathbb{Z}_2^+}\left(\phi^{k,I_{cb}}_{T^2/\mathbb{Z}_2^+}\right)^{\ast} + \int_{{\left|z'\right| \leq \frac{1}{\sqrt{3}}}} dz'd\bar{z'} \ \phi^{i+j+I_{ab}m,I_{cb}}_{S^{2}}\left(\phi^{k,I_{cb}}_{S^{2}}\right)^{\ast}\right] \notag\\
&= \left|\frac{c^{i,I_{ab}}c^{j,I_{ca}}}{c^{k,I_{bc}}}\right| \sum_{m \in \bf{Z}_{I_{bc}}} \vartheta
\begin{bmatrix}
\frac{I_{ca}i-I_{ab}j+I_{ab}I_{ca}m}{-I_{ab}I_{ca}I_{bc}} \\ 0
\end{bmatrix}
\left(0,\tau \left|I_{ab}I_{bc}I_{ca}\right|\right)
f_{i+j+I_{ab}m,k} \notag \\
&\simeq \left|\frac{N_{I_{ab}}N_{I_{ca}}}{N_{I_{bc}}}\right|
\vartheta
\begin{bmatrix}
-\frac{1}{I_{ab}}\left(\frac{j}{I_{ca}}+\frac{k}{I_{bc}}\right) \\ 0
\end{bmatrix}
\left(0,\tau \left|I_{ab}I_{bc}I_{ca}\right|\right). 
\label{eq:Yukawa}
\end{align}
The obtained Yukawa couplings are unchanged after the blow-up, but we note that 
$\left|\frac{\left(\frac{d^{j, I_{ab}}}{c^{j, I_{ab}}}\right)\left(\frac{d^{j, I_{ca}}}{c^{j,I_{ca}}}\right)}{\left(\frac{d_{j, I_{cb}}}{c^{j, I_{cb}}}\right)}\right|=1$ using $I_{ab}+I_{ca}=I_{cb}=-I_{bc}$ and $k=i+j \ {\rm mod} \ I_{ab}$. 
Indeed, when $i,j=0,\frac{M}{2}$, the normalization constant is also the same with the toroidal orbifold one as in Eq.~(\ref{eq:fjk}). 

Similarly, for $0 <j < \frac{M}{2}$, by use of zero-mode product expansion, 
the Yukawa couplings are calculated as
\begin{align}
&Y_{ijk} = \int_{\left|z\right| \geq r} dzd\bar{z} \ \phi^{i,I_{ab}}_{T^2/\mathbb{Z}_2^{+}}\phi^{j,I_{ca}}_{T^2/\mathbb{Z}_2^{+}}\left(\phi^{k,I_{cb}}_{T^2/\mathbb{Z}_2^{+}}\right)^{\ast} + \int_{{\left|z'\right| \leq \frac{1}{\sqrt{3}}}} dz'd\bar{z'} \ \phi^{i,I_{ab}}_{S^2}\phi^{j,I_{ca}}_{S^2}\left(\phi^{k,I_{cb}}_{S^2}\right)^{\ast} \nonumber\\
&= \sum_{m \in \bf{Z}_{I_{bc}}} \frac{1}{\sqrt{2}} \left|\frac{c^{i,I_{ab}}c^{j,I_{ca}}}{c^{i+j+I_{ab}m,I_{cb}}}\right|
\vartheta
\begin{bmatrix}
\frac{I_{ca}i-I_{ab}j+I_{ab}I_{ca}m}{-I_{ab}I_{ca}I_{bc}} \\ 0
\end{bmatrix}
\left(0,\tau \left|I_{ab}I_{bc}I_{ca}\right|\right) \int_{\left|z\right| \geq r} dzd\bar{z} \ \phi^{i+j+I_{ab}m,I_{cb}}_{T^2/\mathbb{Z}_2^{+}}\left(\phi^{k,I_{cb}}_{T^2/\mathbb{Z}_2^{+}}\right)^{\ast} \nonumber \\
&+ \frac{1}{\sqrt{2}} \left|\frac{c^{i,I_{ab}}c^{j,I_{ca}}}{c^{I_{ab}-i+j+I_{ab}m,I_{cb}}}\right|
\vartheta
\begin{bmatrix}
\frac{I_{ca}\left(I_{ab}-i\right)-I_{ab}j+I_{ab}I_{ca}m}{-I_{ab}I_{ca}I_{bc}} \\ 0
\end{bmatrix}
\left(0,\tau \left|I_{ab}I_{bc}I_{ca}\right|\right) \int_{\left|z\right| \geq r} dzd\bar{z} \ \phi^{I_{ab}-i+j+I_{ab}m,I_{cb}}_{T^2/\mathbb{Z}_2^{+}} \left(\phi^{k,I_{cb}}_{T^2/\mathbb{Z}_2^{+}}\right)^{\ast} \nonumber \\
&+ \sqrt{2} \left|\frac{d^{i,I_{ab}}d^{j,I_{ca}}}{d^{k,I_{cb}}}\right|
\vartheta
\begin{bmatrix}
\frac{I_{ca}i-I_{ab}j+I_{ab}I_{ca}m}{-I_{ab}I_{ca}I_{bc}} \\ 0
\end{bmatrix}
\left(0,\tau \left|I_{ab}I_{bc}I_{ca}\right|\right)
\int_{{\left|z'\right| \leq \sqrt{\frac{N-1}{N+1}}}} dz'd\bar{z'} \ \phi^{i+j+I_{ab}m,I_{cb}}_{S^2} \left(\phi^{k,I_{cb}}_{S^2}\right)^{\ast} \nonumber \\
&\simeq \frac{\left|c^{i,I_{ab}}c^{j,I_{ca}}c^{k,I_{cb}}\right|}{\sqrt{2} \left|N_{I_{cb}}\right|^2}
\vartheta
\begin{bmatrix}
-\frac{1}{I_{ab}}\left(\frac{j}{I_{ca}}+\frac{k}{I_{bc}}\right) \\ 0
\end{bmatrix}
\left(0,\tau \left|I_{ab}I_{bc}I_{ca}\right|\right) \notag \\
&- \left(\frac{\pi r^2}{2}\right) \left(\phi^{i,I_{ab}}_{T^2/\mathbb{Z}_2^{+}}\left(0\right)\phi^{j,I_{ca}}_{T^2/\mathbb{Z}_2^{+}}\left(0\right)\left(\phi^{k,I_{cb}}_{T^2/\mathbb{Z}_2^{+}}\left(0\right)\right)^{\ast} + \frac{1}{4\pi} \left(\frac{\pi r^2}{2}\right) \left[\phi^{i,I_{ab}}_{T^2/\mathbb{Z}_2^{+}}\left(0\right)\phi^{j,I_{ca}}_{T^2/\mathbb{Z}_2^{+}}\left(0\right)\left(\phi^{k,I_{cb}}_{T^2/\mathbb{Z}_2^{+}}\left(0\right)\right)^{\ast}\right]'' \right) \nonumber \\
&+ 2\sqrt{2} \left(\frac{\pi r^2}{2}\right) \left(\phi^{i,I_{ab}}_{T^2}\left(0\right)\phi^{j,I_{ca}}_{T^2}\left(0\right)\left(\phi^{k,I_{cb}}_{T^2}\left(0\right)\right)^{\ast} + \frac{1}{4\pi} \left(\frac{\pi r^2}{2}\right) \left[\phi^{i,I_{ab}}_{T^2}\left(0\right)\phi^{j,I_{ca}}_{T^2}\left(0\right)\left(\phi^{k,I_{cb}}_{T^2}\left(0\right)\right)^{\ast}\right]'' \right) \nonumber \\
&= \frac{\left|c^{i,I_{ab}}c^{j,I_{ca}}c^{k,I_{cb}}\right|}{\sqrt{2} \left|N_{I_{cb}}\right|^2}
\vartheta
\begin{bmatrix}
-\frac{1}{I_{ab}}\left(\frac{j}{I_{ca}}+\frac{k}{I_{bc}}\right) \\ 0
\end{bmatrix}
\left(0,\tau \left|I_{ab}I_{bc}I_{ca}\right|\right) \nonumber \\
&+ \frac{1}{4\pi} \left(\frac{\pi r^2}{2}\right)^2 \left( 2\sqrt{2} \left[\phi^{i,I_{ab}}_{T^2}\left(0\right)\phi^{j,I_{ca}}_{T^2}\left(0\right)\left(\phi^{k,I_{cb}}_{T^2}\left(0\right)\right)^{\ast}\right]'' - \left[\phi^{i,I_{ab}}_{T^2/\mathbb{Z}_2^{+}}\left(0\right)\phi^{j,I_{ca}}_{T^2/\mathbb{Z}_2^{+}}\left(0\right)\left(\phi^{k,I_{cb}}_{T^2/\mathbb{Z}_2^{+}}\left(0\right)\right)^{\ast}\right]''\right). 
\label{eq:Yukawa2}
\end{align}
The first term is coming from $k=i+j \ {\rm mod} \ I_{ab}$ or $k=I_{ab}-i+j \ {\rm mod} \ I_{ab}$.
When $k=0, \frac{I_{cb}}{2}$, the right-handed side becomes $\sqrt{2}$ times Eq.~(\ref{eq:Yukawa2}) and the above 
Yukawa couplings are the same with Eq.~(\ref{eq:Yukawa}) when $i,j =0, \frac{I}{2}$.

\end{document}